\documentclass[9pt,twocolumn,twoside]{osajnl}

\journal{ol} 

\setboolean{shortarticle}{true}

\usepackage{lineno}

\title{Monolithically integrated circuits for Optical injection locking of ring lasers with QKD and QPSK applications.}

\author[1,*]{Damiano Massella}
\author[2]{Michael Wallace}
\author[2]{Ronald Broeke}
\author[1,3]{Francisco Diaz}
\affil[1]{University of Vigo, El Telecomunication - Campus Universitario As Lagoas, 36310 Vigo, Spain}
\affil[2]{Bright Photonics BV, Eindhoven, Netherlands.}
\affil[3]{AtlanTTic research center, El Telecomunication - Campus Universitario As Lagoas, 36310 Vigo, Spain}

\affil[*]{Corresponding author: damiano.massella@uvigo.es}

\begin{abstract}

Optical injection locking of integrated ring lasers promises to be a key technology of the future, but it is still lacking experimental verification. In this paper we present two monolithically integrated photonic circuits that  experimentally realize  Optically injection locking of ring lasers. The presented circuits pave the way for future application in Quantum Key Distribution and coherent communication. 

\end{abstract}

\setboolean{displaycopyright}{true}

\begin{document}

\maketitle

\section{Introduction}

In recent years the discussion regarding communication security has taken a central role as the importance of telecommunication grows. Quantum Key Distribution (QKD) promises to solve the problem of security in communication, an issue which is given more significance with the advent of quantum computers\cite{qiu_quantum_2014}. \\
At the same time, due to ever increasing demands on data transmission there has been growing interest in Optical Injection Locking (OIL) of lasers\cite{liu_scaling_2010}. Using this technique it is possible to increase the modulation bandwidth of a laser up to 100GHz\cite{liu_optical_2020}. \\
The bandwidth of transmitters can be increased by multiplexing a large number of channels however, this is not scalable using bulk optics. Photonic Integrated Circuits (PICs), consist of a large number of optical components integrated on a single chip, are a key technology to overcome this limitation \cite{tanaka_high-speed_2012}.\\
QKD transmitters have been realized in both Indium Phosphide and Silicon showing the advantage of integration in terms of stability and miniaturization of single channels.
The work of Sibson et al  \cite{sibson_chip-based_2017} reports on the integration of a QKD transmitter without the use of modulators and based on an OIL technique first demonstrated in \cite{yuan_directly_2016}. 
In this paper we report the first experimental realization OIL of monolithically integrated ring laser systems. 
An extensive study of this kind of system has been reported before but no experimental realization has been show yet \cite{zhang_modulation_2022,duzgol_modulation_2017,chrostowski_monolithic_2008}.

In this work we show two different circuits designed for two different applications. The first circuit is consist of a Distributed Feedback (DFB) Laser is OIL to a ring laser. This system is intended for mainly QKD purposes. Coherent communication is also possible however, the modulation scheme would be too complex\cite{paraiso_modulator-free_2019}. \\
Our second circuit is realized using 2 ring lasers that are feed using a Distributed Bragg Reflector (DBR) laser. This circuit has been designed to be a coherent communication transmitter and thus be able to modulate both the ring lasers simultaneously. \newline
With respect to the circuit presented by Paraiso et all \cite{paraiso_modulator-free_2019}, the use of ring lasers instead of DFB lasers carries some significant advantages.\newline
Firstly, the ring lasers don't need to be tuned like in the case of single mode lasers. 
And secondly, when OIL a ring laser we ensure the unidirectional emission from it. Consequently, there are no reflections from the slave laser to the master laser, like in the DFB case \cite{liu_optical_2020}.\newline




\section{Circuit Design}

\begin{figure}[h!]

    \centering
    \includegraphics[trim=0cm 0cm 0cm 0cm,, clip=false,width=\linewidth]{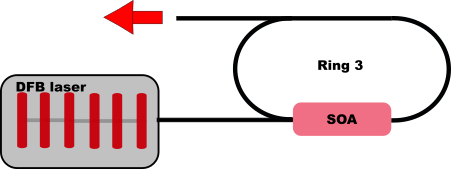}
    \caption{This schematic represent  the first design of OIL ring laser. The ring laser is injected using a DFB laser. }
    \label{fig:dfb_scheme}
\end{figure}

The first circuit is presented in figure \ref{fig:dfb_scheme}.
In this circuit we use a simple DFB laser to inject a ring laser.
The notation for the ring lasers follows the one used for the mask design, consequently in the ring laser of circuit one is labelled as 'ring 3'.
The simplicity of this circuit is its main advantage with respect to other QKD circuits.
In the schematics of this section we have omitted the MMI blocks and the photodiodes used to monitor the state of the laser.\\

We used a standard DFB laser designed to emit at $1550nm$, the left output of it is terminated using a photodiode.
This photodiode will be used to characterize the DFB laser emitted power.\\
On the ring we have used custom MMI blocks, they are designed to split power unequally, $85\%$ on the cross state and $15\%$ on the bar.
We have chosen this particular ratio to reduce the losses inside the ring and have a shorter gain section.
The MMI used in the rings are all $2x2$ and the additional port is terminated using a photodiode.
The photodiodes inside the ring can be used to monitor both the clockwise directed power and the counter-clockwise.
The SOA section is $400 \mu m$  long and it has been connected using RF lines in order to be able to drive the ring laser to higher frequencies and verify the modulation bandwidth enhancement due to OIL.
A spot size converter is placed at the output of the circuit to minimize the losses when coupling to fiber.\\

\begin{figure}[h!]

    \centering
    \includegraphics[trim=0cm 0cm 0cm 0cm,, clip=false,width=\linewidth]{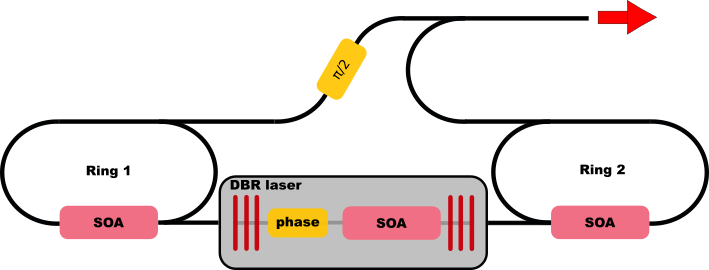}
    \caption{This schematic represents the second design of OIL ring laser. The two ring lasers are injected using a DBR laser.}
    \label{fig:dbr_scheme}
\end{figure}

The second circuit is schematized in figure \ref{fig:dbr_scheme}.\\
In this case we decided to exploit the two outputs of a DBR laser to OIL two ring lasers. The output of the two rings are then combined to a single output. This schematic can be used for both QKD and Quadrature Phase Shift Keying (QPSK).\\
The main advantage with respect to the first circuit is the possibility to use the modulation of the rings for encoding the bits of the QPSK scheme.
We have added a phase modulator in the output of one of the ring laser in order to have the $\pi/2$ phase shift for QPSK.
This additional phase modulation section is constituted by an thermo-optical phase modulator of length $450\mu m$.
This length is sufficient to ensure the $\pi/2$ phase difference between the two rings.
The modulation bandwidth is enhanced by the OIL since we are going to modulate the two slave lasers and not the master laser.\\
The ring lasers used for the two circuits are identical in dimensions and components used.
The constituents components of the DBR laser have lengths of: $30\mu m$ DBR mirrors, $400\mu m$ SOA and $200\mu m$ phase modulation section.

The circuits presented in this letter are fitted in an MPW cell of dimensions $8mm\: x\: 2mm$, highlighting the small footprint of both of them respect to traditional modulator based circuits.

\section{Measurements and discussion}

In the following section we are going to present the experimental results on the basic functionalities of the circuit.

\begin{figure}[h!]

    \centering
    \includegraphics[trim=0cm 0cm 0cm 0cm,, clip=false,width=\linewidth]{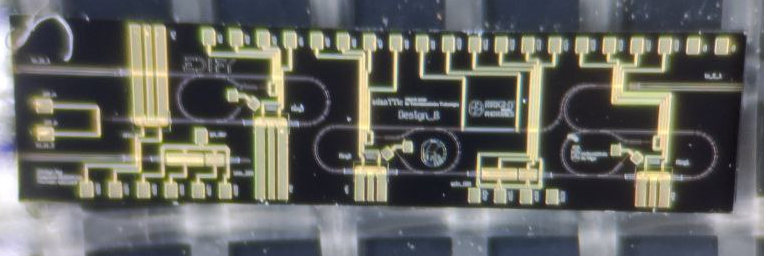}
    \caption{Photo of the realized circuit under a microscope. We can
        clearly see the metal tracks for electrical connection and
        the waveguides below them.}\label{fig:designB_photo}
\end{figure}

\subsection{DFB and single ring circuit}

When analyzing the emission of the DFB laser it became immediately evident that there had been some problem in the fabrication. In fact, the threshold is above $60mA$ against the $25mA$ expected from the design manual.
This has been confirmed from the fabrication report and in general we can observe a lot of variability in the behaviour of this component.
We have measured other DFB laser fabricated in the same run and the threshold current ranges from $44mA$ to $60mA$ with maximum emitted power above $1mW$ only in one case.\newline   

To estimate the MMI performances we drove the DFB at $80mA$, resulting in a current at the photodiode terminating the laser of $-80\mu A$ whereas, the photodiode in the cross port of the MMI registered $-60\mu A$. Both the photodiodes were biased at $-2V$ during the measurement.\\
Assuming a $1dB$ loss for this kind of component we reach the result that the splitting ratio of the MMI is $83\%$ for the cross port and $17\%$ for the through port. This ratio is close to the designed ratio of $85-15$.

\begin{figure}[h!]

    \centering
    \includegraphics[trim=0cm 0cm 0cm 0cm,, clip=false,width=\linewidth]{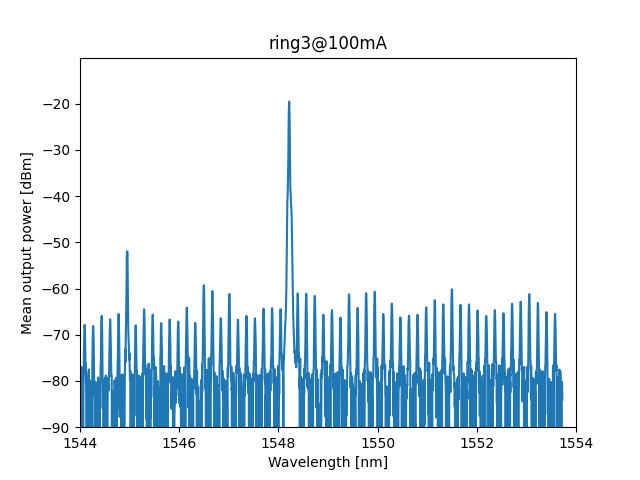}
    \caption{Emission spectra of the ring 3 as measured in the lab. The single emission peak is due to cross-gain saturation and to fabrication defects. We can notice the frequency comb in the low power regime.}
    \label{fig:ring3_exp_spectra}
\end{figure}

Before we discuss the OIL of the ring laser we have to address it's free running properties.
Figure \ref{fig:ring3_exp_spectra} reports the experimental spectrum recorded by coupling a fiber to the chip and driving the ring3 SOA at $100mA$.
We can notice that in this case the laser is single mode with lots of small peaks representing the cavity resonances.
The laser emission is not stable and tends to jump and side modes appear and disappear following non-predictable fluctuations.
This is due to the intrinsic multi-mode nature of the laser.\\

\begin{figure}[h!]

    \centering
    \includegraphics[trim=0cm 0cm 0cm 0cm,, clip=false,width=\linewidth]{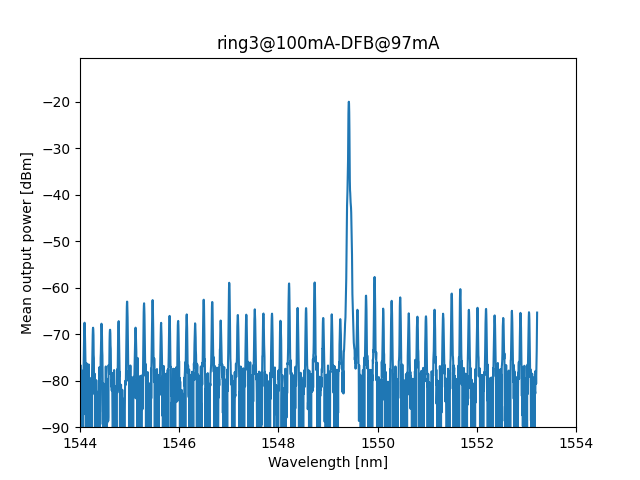}
    \caption{Spectrum of the ring3 laser at $100mA$ when it is injected with the DFB at $97mA$. We can notice the shift in wavelength with respect to figure \ref{fig:ring3_exp_spectra} that is due to the optical injection of the DFB.}
    \label{fig:ring3_dfb_spectra}
\end{figure}

The next step of the experimental analysis is to verify the OIL of the system.\\
Figure \ref{fig:ring3_dfb_spectra} reports the resulting spectra from optical injection of the ring laser. The spectra is obtained applying a current of $100mA$ to the ring3 SOA and $97mA$ to the DFB. The result proves the OIL, since we do not see any other peak and the main peak is shifted respect to figure \ref{fig:ring3_exp_spectra}. Comparing the two figures we can also observe that the secondary peak seen in figure \ref{fig:ring3_dfb_spectra} at $1545nm$ disappears. \\

An additional proof of OIL is obtained by comparing the photodiodes recorded power. In fact, the optical injection of a ring laser ensures the unidirectionality of the light inside the cavity, completely suppressing the CW mode. In this case, we have measured the current at the photodiode in the CW direction and turned on and off the DFB laser. Without DFB laser injection the ring laser is bidirectional: the current on the CW photodiode is $-0.160mA$. When the DFB laser is turned on at $97mA$, the current measured on the CW photodiode is $-0.002mA$, demonstrating the high suppression of the CW light and the successful OIL.

\subsection{DBR and double ring circuit}

We will start our analysis from ring2, this ring is identical to ring1 and really similar to ring3 analyzed in the previous section.
The main difference with ring3 is the presence, in ring1 and ring2, of the feedback of the DBR mirror of the master laser. This changes the behaviour of the ring, giving it a preferred lasing direction.\\

\begin{figure}[h!]

    \centering
    \includegraphics[trim=0cm 0cm 0cm 0cm,, clip=false,width=\linewidth]{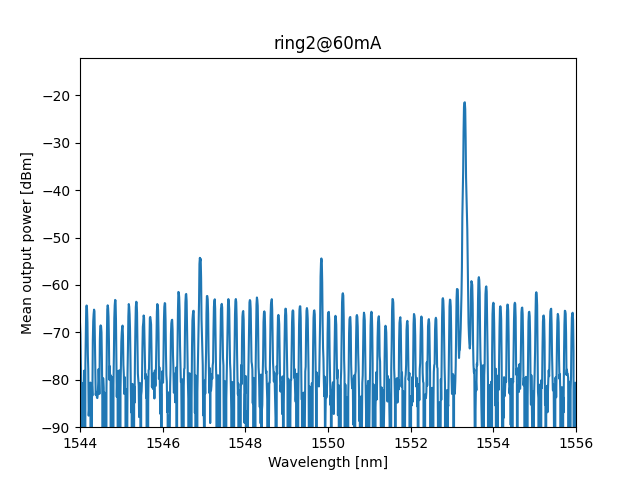}
    \caption{Emission spectra measured at $60mA$ current injection on the ring2 $SOA$. As in the previous case the ring exhibits single mode operation due to small imperfections in the fabrication and cross-gain saturation.}
    \label{fig:ring2_exp}
\end{figure}

The spectrum of Ring2 is presented in figure \ref{fig:ring2_exp}.\\
Similarly, to the previous ring laser we notice a single peak. This can be account by
small reflections due to fabrication defects can lead to different cavity length and ultimately to single mode lasing.
A second effect not taken into account is the cross-gain saturation effect that tends to make only a single wavelength lase.\\
It is typical of a multimode laser to change lasing frequency based on small changes in temperature and conditions of the cavity, leading to a highly unstable emission.

The design of the DBR laser was focused on having a high power output sacrificing the single mode emission. In fact, we will use the OIL not only for isolating a single frequency of the slave laser but also to isolate a single frequency of the master laser. The challenging part will then be to lock both ring1 and ring2 laser to the same master frequency.\\

\begin{figure}[h!]

    \centering
    \includegraphics[trim=0cm 0cm 0cm 0cm,, clip=false,width=\linewidth]{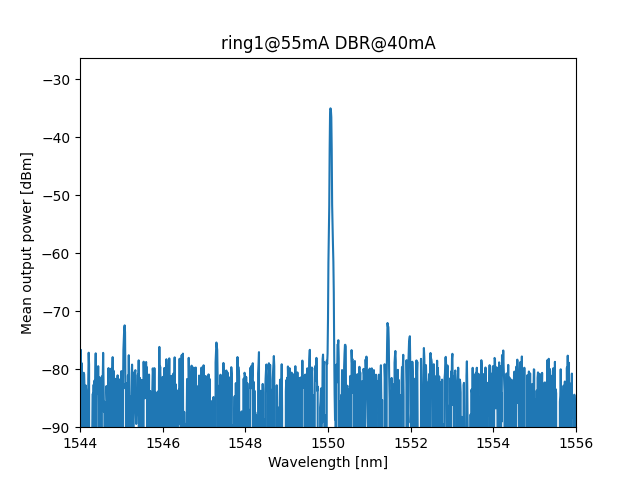}
    \caption{Emission spectra of ring1 at $55mA$ when injected with the DBR laser at $40mA$. We can notice the single mode emission resulting from the locking of the DBR and the ring.}
    \label{fig:ring1_injected}
\end{figure}

We can now discuss the OIL scheme. We first prove the locking with only one of the rings.\\
Figure \ref{fig:ring1_injected} shows the output of ring1 laser when injected with the DBR laser.\\
The laser emission is single mode with a side-mode suppression ratio of $45dB$ demonstrating the locking of the two lasers. In fact, we do not see the multimode emission of the DBR laser nor any sign of the multimode of the ring laser. This shows that this system can be used for OIL effectively.\\
Like in the case of the DFB and ring3 the current applied to the SOA is critical and influences the locking mechanism, but we can find multiple combinations of currents that achieve OIL.\\
This system is already capable of QKD like the previous one, but we also want to test if we are able to lock both ring1 and ring2 to the same DBR laser mode at the same time.\\

When the ring2 laser is activated we notice that there is a series of other peaks in addition to the peak from ring1 in the spectrum.
We have varied the current applied to the different lasers in order to get to stable locking of both ring lasers, but the best results so far has been not convincing.
These results are clearly not usable for QPSK since for this application we need a single wavelength.
The main question raising from this plot is why the injection doesn't work for both rings ?

\begin{figure}[h!]

    \centering
    \includegraphics[trim=0cm 0cm 0cm 0cm,, clip=false,width=\linewidth]{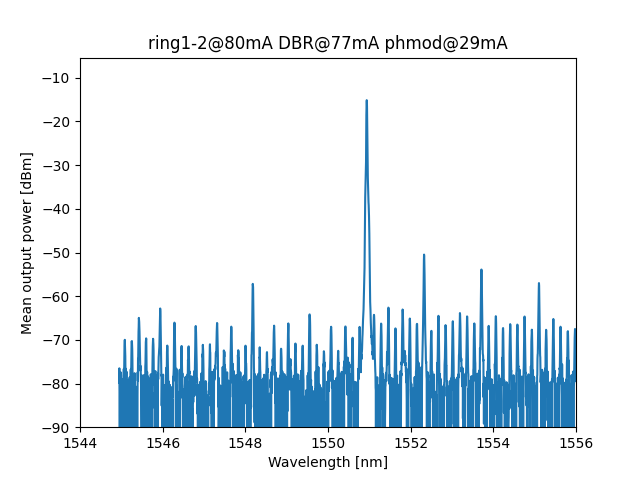}
    \caption{Spetrum, ring1 and ring2 are injected at $80mA$, the SOA of the DBR laser is driven at $77mA$ and the phase modulation section of the DBR laser is driven at $29mA$. This combination of factors has been chosen to realize optical injection in both the rings. }
    \label{fig:rings_injection_ph}
\end{figure}

As a final test we have decided to vary the phase modulator in the DBR laser in order to check the influence of it in the locking of the lasers.\\
We have decided to drive the rings at the same current of $80mA$ and only vary current applied to the DBR laser SOA section and phase section.
The best recorded spectra is visible in figure \ref{fig:rings_injection_ph}, obtained with $77mA$ on the SOA and $29mA$ on the phase modulator.
The resulting spectrum shows a main peak and a few side peaks $35dB$ lower.
From this spectrum is still clear that the system is still not necessary locked and the three laser influence each other.
In fact if we turn ring1 off the output spectra goes back to be multimode.
This gives the indication that the rings themselves are influencing the emission of one another, even if in an OIL system they should be lasing in a single direction as proven for ring3. \\
Further investigation in this system is necessary and the photodiodes inside ring1 and ring2 should be monitored to understand if there is light travelling in the counter propagating mode of the laser.

\section{Conclusions}
In conclusion, we have designed two photonic integrated circuits for QKD and QPSK both based on OIL mechanism.\\
The first circuit presented is composed by a DFB and a ring laser. We have characterized the laser's free running behaviour. Using the integrated photodiodes and an optical spectrum analyzer we have demonstrated OIL between the two lasers, resulting in single mode, single direction emission.
Using the DFB and the two photodiodes it has been possible to estimate the power splitting ratio of the custom MMI, in agreement with the designed specifications.
This is the first experimental verification of monolithically integrated OIL system using ring lasers.\\
The second circuit is composed of two ring lasers and a central DBR laser. Using OIL we want to achieve high speed modulation of the ring lasers simultaneously. We have characterized the identical ring lasers and DBR laser separately. We have demonstrated the OIL of one ring laser with the DBR laser, but when moving to locking both ring lasers to the master DBR laser we achieved partial locking only using the phase tuning section.\\
In future work we want to measure the modulation bandwidth of both circuits and further investigate the double ring locking system.
\section{Acknowledgements}

Project developed in the framework of the 
European Doctorate in Indium Phosphide PIC Fabrication Technology (EDIFY) project.

\section{Disclosures}
The authors declare no conflicts of interest.

\bibliography{30_references/QKD_paper.bib}

\bibliographyfullrefs{30_references/QKD_paper.bib}


\ifthenelse{\equal{\journalref}{aop}}{%
\section*{Author Biographies}
\begingroup
\setlength\intextsep{0pt}
\begin{minipage}[t][6.3cm][t]{1.0\linewidth} 
  \begin{wrapfigure}{L}{0.25\linewidth}
    \includegraphics[width=0.25\linewidth]{john_smith.eps}
  \end{wrapfigure}
  \noindent
  {\bfseries John Smith} received his BSc (Mathematics) in 2000 from The University of Maryland. His research interests include lasers and optics.
\end{minipage}
\begin{minipage}{1.0\linewidth}
  \begin{wrapfigure}{L}{0.25\linewidth}
    \includegraphics[width=0.25\linewidth]{alice_smith.eps}
  \end{wrapfigure}
  \noindent
  {\bfseries Alice Smith} also received her BSc (Mathematics) in 2000 from The University of Maryland. Her research interests also include lasers and optics.
\end{minipage}
\endgroup
}{}

\end{document}